\documentclass[prb,aps,twocolumn,superscriptaddress,showpacs,eqsecnum]{revtex4}
\usepackage{epsfig}
\usepackage{amsmath}
\usepackage{amsthm}
\usepackage{amssymb,braket}
\usepackage{verbatim}
\usepackage{bm}
\usepackage{hyperref}
\usepackage[all]{hypcap}
\usepackage{here}
\usepackage{sidecap}
\usepackage{float}
\usepackage{graphicx}
\usepackage{wasysym}
\usepackage{color}
\usepackage{amssymb}
\usepackage{siunitx}
\usepackage{ifpdf}

\newcommand{\be}{\begin{eqnarray}}
\newcommand{\ee}{\end{eqnarray}}
\newcommand{\bbm}{\begin{bmatrix}}
\newcommand{\ebm}{\end{bmatrix}}
\newcommand{\bpm}{\begin{pmatrix}}
\newcommand{\epm}{\end{pmatrix}}
\renewcommand{\v}[1]{{\bf #1}}
\newcommand{\nn}{\nonumber \\}

\newcommand{\angstrom}{\text{\normalfont\AA}}

\begin{document}
\title{Experimental determination of the massive Dirac fermion model parameters for MoS$_2$, MoSe$_2$, WS$_2$, and WSe$_2$}

\author{Beom Seo Kim}
\affiliation{Center for Correlated Electron Systems, Institute for Basic Science, Seoul 151-747, Korea}
\affiliation{Department of Physics and Astronomy, Seoul National University, Seoul 151-747, Korea}
\affiliation{Department of Physics, Incheon National University, Incheon 406-772, Korea}

\author{Jun-Won Rhim}
\email[Electronic address for computational part: $~~$]{phyruth@gmail.com}
\affiliation{Max-Planck-Institut f\"{u}r Physik komplexer Systeme, 01187 Dresden, Germany}

\author{Beomyoung Kim}
\affiliation{Department of Physics, Pohang University of Science and Technology, Pohang 790-784, Korea}
\affiliation{Advanced Light Source, Lawrence Berkeley National Laboratory, Berkeley, CA 94720, USA}

\author{Changyoung Kim}
\affiliation{Center for Correlated Electron Systems, Institute for Basic Science, Seoul 151-747, Korea}
\affiliation{Department of Physics and Astronomy, Seoul National University, Seoul 151-747, Korea}

\author{Seung Ryong Park}
\email[Electronic address for experimental part: $~~$]{AbePark@incheon.ac.kr}
\affiliation{Department of Physics, Incheon National University, Incheon 406-772, Korea}

\begin{abstract}
Monolayer MX$_2$ (M = Mo, W; X = S, Se) has drawn much attention recently for its possible application possibilities for optoelectronics, spintronics, and valleytronics. Its exotic optical and electronic properties include a direct band gap, circular polarization dependent optical transitions, and valence band (VB) spin band splitting at the $K$ and $-K$ points. These properties can be described within a minimal model, called the massive Dirac fermion model for which the parameters need to be experimentally determined. We propose that the parameters can be obtained from angle resolved photoemission (ARPES) data from bulk 2H-MX$_2$, instead of monolayer MX$_2$. Through tight binding calculations, we show how the electronic structure at high symmetry points evolves as the system changes from the monolayer to the three dimensional bulk 2H-MX$_2$ . We find vanishing $k_z$ dispersion and almost no change in the direct band gap at the $K$ and $-K$ points, in sharp contrast to the strong $k_z$ dispersion at the $\Gamma$ point. These facts allow us to extract the gap and spin band splitting at the $K$ point as well as the hopping energy from bulk ARPES data. We performed ARPES experiments on single crystals of MoS$_2$, MoSe$_2$, WS$_2$, and WSe$_2$ at various photon energies and also with potassium evaporation. From the data, we determined the parameters for the massive Dirac fermion model for monolayer MoS$_2$, MoSe$_2$, WS$_2$, and WSe$_2$.
\end{abstract}

\maketitle

\section{Introduction}

The successful exfoliation of graphene\cite{Novoselov,Geim,Kim} is important on its own right but also has triggered the intensive/extensive research on similar two-dimensional layered materials. Transition metal dichalcogenides (TMDs) such as NbSe$_2$ and MoS$_2$ have strong in-plane covalent and weak out-of-plane van der Waals bonds, which reduce the dimensionality from three to two and allow us to obtain monolayer systems by the exfoliation method. Monolayer TMDs often exhibit qualitatively different electronic properties compared to the bulk\cite{Mak,Mo,Chhowalla}.

Among the TMDs, the group 6 TMDs MX$_2$ (M = Mo, W; X = S, Se) exhibit interesting electronic properties such as indirect to direct band gap transition from bulk to monolayer \cite{Mak,Mo}, valley degeneracy\cite{Xiao}, and spin-orbit interaction (SOI) induced spin band splitting at the $K$ and $-K$ points of the hexagonal Brillouin zone\cite{Zhu}. From these fundamental electronic properties, the valley degeneracy can be lifted by using circularly polarized light\cite{Cao,Zeng,Heinz,Aivazian,Hsu} and valley Hall effect was observed\cite{McEuen,Tahir,Olsen}. These raised the notion of the valleytronics\cite{Yuan,JHKim,Sie,Wang,C.Zhu,Li,MacNeill}.

It would be desired to have a simple model that covers these exotic properties of group 6 TMDs for practical purpose. A minimal model, massive Dirac fermion model, is simple but can cover all the interesting low energy electronic structure properties mentioned above\cite{Xiao}. The model has only three independent parameters: the effective hopping ($t$), band gap without SOI ($\Delta$), and spin band splitting ($2\lambda$). The details of the model are described in III-A. The electronic structure of TMDs can be directly measured by ARPES, which has confirmed the direct band gap and the spin band splitting at the $K$ and $-K$ points\cite{Mo,Riley,Jin,Latzke,King,Yeh,W.Jin,Miwa,Dendzik}. More importantly, the parameters for the massive Dirac fermion model can be directly measured by using ARPES. The measured values of $\Delta$ and $2\lambda$ are $1.465$ and $0.15$ eV for the epitaxial monolayer MoS$_2$ on Au($111$), and $1.67$ and $0.18$ eV for monolayer MoSe$_2$ grown on graphene\cite{Mo,Miwa}.

There are a couple of obstacles in experimentally measuring the massive Dirac fermion model parameters by ARPES. The experiments have been mostly performed on epitaxially grown MX$_2$ monolayer systems due to the difficulty in ARPES experiments on exfoliated MX$_2$ monolayer. Epitaxial strain and formation of superstructure due to the interaction with the substrate may affect the parameters whereas other experiments such as transport measurements have been mostly done on exfoliated MX$_2$ monolayer systems. It is now possible to do ARPES on exfoliated MX$_2$ monolayer with the size of tens of micrometers with the development of so-called micro-ARPES for which the incident light is focused to sub micrometer size. Unfortunately, however, the quality of the data from exfoliated MX$_2$ monolayer by micro-ARPES is still not good enough to extract the parameters quantitatively\cite{Jin,Yeh,W.Jin}.

Our idea is to extract the parameters from ARPES data from bulk systems instead of monolayer MX$_2$. Even though the massive Dirac fermion model is applicable only for monolayer MX$_2$, we show that we can extract the massive Dirac fermion parameters from the electronic structure of bulk materials. Tight binding calculation result shows how the electronic structure at high symmetry points evolves as the system changes from the two dimensional monolayer to the three dimensional bulk 2H-MX$_2$. It also reveals that the direct band gap at $K$ and $-K$ points for the monolayer is identical to that for the bulk due to lack of the $k_z$ dispersion. Our photon energy dependent ARPES data indeed confirms the vanishing $k_z$ dispersion near the $K$ and $-K$ points. Accordingly, all the appropriate parameters ($t$, $\Delta$ and $2\lambda$) for MoS$_2$, MoSe$_2$, WS$_2$, and WSe$_2$ systems were successfully extracted from the experimental ARPES data. From now on, we omit "bulk 2H-" for the bulk sample.

\begin{figure}
\centering \epsfxsize=8.5cm \epsfbox{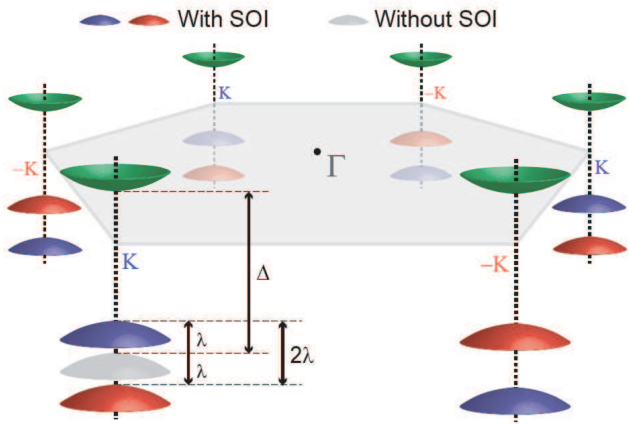}
\caption{Schematic sketch of the massive Dirac fermion model. Gray VB at the front-left $K$ point is for the case without SOI while red/blue VB edges correspond to the spin up/down states for the case with SOI.}\label{fig1}
\end{figure}

\section{Methods}
ARPES measurements were done at the beam line 4.0.3.2 (MERLIN) of the Advanced Light Source equipped with a VG-SCIENTA R8000 analyzer. The total energy resolution was better than $20$ meV. Four high quality single-crystal samples were purchased from 2D Semiconductors and HQGraphene. All the data were taken under $40$ K in a base pressure better than $4.5\times 10^{-11}$ Torr. For the photon energy dependence, we used the photon energy between $50$ and $100$ eV. Alkali Metal Dispensers from SAES Getters were used for potassium evaporation experiments and evaporation was conducted $\emph{in situ}$ with the samples at the measurement position.

\section{Results and Discussion}
\subsection{Tight binding calculations for electronic-structure evolution from monolayer to bulk MX$_2$}

Figure 1 is a schematic sketch of the massive Dirac-fermion model. Two cases are illustrated in the figure, one without SOI and the other with SOI. The Hamiltonian of the massive Dirac Fermion model including SOI reads

\be
\hat{H} = at(\tau{k_{x}}{\hat{\sigma}_{x}} + {k_{y}}{\hat{\sigma}_{y}}) + \frac{\Delta}{2}{\hat{\sigma}_{z}} - \lambda\tau{\frac{{\hat{\sigma}_{z}} - 1}{2}{\hat{s}_{z}}}
\ee
where $a$ is the lattice constant, $t$ the effective hopping parameter, $\tau$ the valley index, $\hat{\sigma}$ the Pauli matrices for the basis functions, $\Delta$ the direct band gap size without SOI, 2$\lambda$ the SOI induced spin band splitting size, and $\hat{s}_{z}$ the Pauli matrix for spin (see Ref.[7] for more details). Note that there are only three free-parameters in this model, $\Delta$, 2$\lambda$, and $t$. As stated earlier, the goal of our research is to determine these parameters experimentally for MoS$_2$, MoSe$_2$, WS$_2$, and WSe$_2$ monolayer. On the other hand, ARPES experiments were performed on bulk MX$_2$ for which the low-energy electronic properties are not governed by the massive Dirac Fermion model. For example, the VB maximum is not located at the $K$ point but at the $\Gamma$ point in MX$_2$. We performed tight binding calculations with a focus on how the electronic structure at the $K$ and $\Gamma$ points evolves from monolayer MX$_2$ to bulk MX$_2$. Our calculations show that electronic structure evolution at $K$ point is small enough that we can extract the massive Dirac fermion parameters from the electronic structure of MX$_2$.

First, we consider the $k_z$ dependent VB dispersion at the in-plane $\Gamma$ point. The conduction band (CB) is not treated here since it is not easy to find an effective model for this band due to the multiple mixing with other bands. Meanwhile, the VB is well separated from other bands and the mixing could be negligible. In this paper, we neglect the spin degree of the freedom which does not affect the band broadening. At $\Gamma$, the orbital composition of the VB is known to be
\be
|\psi_{\Gamma_{VB}}\rangle &=& \tilde{c}_1 |d_0^{(e)}\rangle - c_1 |p_0^{(e)}\rangle
\ee
where $|d_0^{(e)}\rangle = | d_{z^2}\rangle$ and $|p_0^{(e)}\rangle = ( |p_z^{A}\rangle  - |p_z^{B}\rangle )/\sqrt{2}$. Here, $A$ and $B$ represent the chalcogen atoms at the upper and lower side of the MX$_2$ slab. From now on, we omit the superscript $(e)$ of the $p_z$ orbital. $\tilde{c}_1 = \sqrt{1-c_1^2}$ and its value for various TMDs has been obtained by Fang \textit{et. al.}.\cite{shiang}

One can construct a Bloch wave function with the translational symmetry along $z$-axis as
\be
|\Psi_{\Gamma_{VB},k_z}^{l(u)}\rangle &=& \frac{1}{\sqrt{N}} \sum_{n} |\psi_{\Gamma_{VB}}^{n,l(u)}\rangle e^{in k_z c}
\ee
where $n$ is the layer index, and $l$ and $u$ represent lower and upper MX$_2$ slab in the unit cell. $c$ is the lattice constant along $z$ direction and we set the gauge so that there is no $k_z$ dependence in the same unit cell. Here, $|\psi_{\Gamma_{VB}}^{n,l(u)}\rangle$ is a function of $k_x$ and $k_y$ and is constructed to satisfy the Bloch condition in the $xy$-plane.

\begin{figure}
\centering \epsfxsize=8.5cm \epsfbox{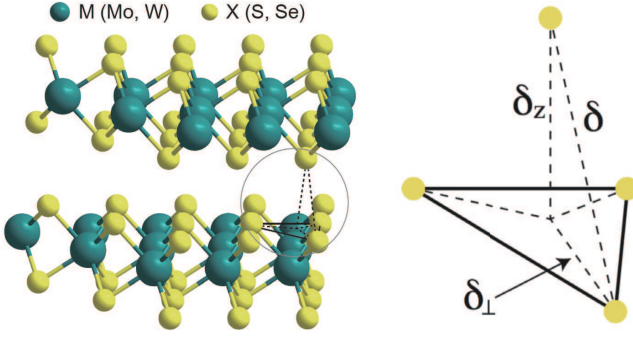}
\caption{Crystal structure of MX$_2$. Inversion symmetry is broken within an MX$_2$ monolayer. In MX$_2$, the layers are stacked in a way that inversion symmetry is restored. The figure on the right is the zoom-in of the part marked by the circle in the crystal structure. It shows the definitions of the parameters used in the calculation.}\label{fig2}
\end{figure}

If we assume that $\epsilon_{\Gamma_{VB}}$ is the VB energy of MX$_2$ monolayer, the effective Hamiltonian for the 3D bulk system at the point is given by
\be
H_{\Gamma} \approx \bpm \epsilon_{\Gamma_{VB}} & \Delta_{\Gamma,k_z} \\ \Delta_{\Gamma,k_z}^* & \epsilon_{\Gamma_{VB}} \epm \quad
\ee
where $\Delta_{\Gamma,k_z} = \langle \Psi_{\Gamma_{VB},k_z}^{u}| H^\prime | \Psi_{\Gamma_{VB},k_z}^{l}\rangle $.
Here, $H^\prime$ is the interlayer hopping term which will be explained below. Its eigenvalues are evaluated to be
\be
E_{\Gamma_{VB}}^{\pm} &=&\epsilon_{\Gamma_{VB}} \pm |\Delta_{\Gamma,k_z}|
\ee
The off-diagonal component $\Delta_{\Gamma,k_z}$, which is dependent on $k_z$, corresponds to the band broadening and can be calculated as follows.
\be
\Delta_{\Gamma,k_z} &\approx & -\frac{c_1^2}{2} \sum_i \langle p^{B,n,u}_z | H^\prime  |p^{A,n,l}_z(\delta_i)  \rangle \nn
&& -\frac{c_1^2}{2} \sum_i \langle p^{A,n,u}_z | H^\prime |p^{B,n+1,l}_z(\tilde{\delta}_i)\rangle e^{ik_z c}\label{eq:gamma_4}
\ee
where the vector $\delta_i$ represents the nearest neighbor sites between MX$_2$ layers, and $\tilde{\delta}_i = -\delta_i$. Note that the nearest neighbor vectors between slabs in the same unit cell are in opposite direction to those in different unit cells due to the way of the stacking.\cite{xiangying} There are no phase factors that are dependent on $k_x$ and $k_y$ in the above since we are considering only the $\Gamma$ point. Details are presented in Appendix A.

Now, we use the following Slater-Koster approximation.
\be
t^{(LL)}_{p^\prime_i,p_j} &=& \langle p^\prime_i(\v r_i) | H^\prime | p_j(\v r_j) \rangle \\
&=& \left( V_{pp\sigma} - V_{pp\pi}\right)\frac{r_i r_j}{r^2} + V_{pp\pi}\delta_{ij} \label{eq:slater_koster}
\ee
where $V_{pp\sigma(\pi)}$ is an exponentially decaying function of the distance between $p$ orbitals.\cite{shiang} For the case of $\Delta_{\Gamma,k_z}$, only the $p_z$ orbital is involved, so that $r_ir_j/r^2 = (\delta_z/\delta)^2$ (see Fig. 2). As a result, we have
\be
\Delta_{\Gamma,k_z} &=& -\frac{c_1^2}{2}\sum_i\langle p^{B,n,u}_z | H^\prime |p^{A,n,l}_z(\delta_i)\rangle \left(1+e^{ik_zc}\right)  \nn
&=& -\frac{D_\Gamma}{2} \left(1+e^{ik_zc}\right)
\ee
where
\be
D_\Gamma = 3 c_1^2 \left\{ \left( V_{pp\sigma} - V_{pp\pi}\right)\left(\frac{\delta_z}{\delta}\right)^2 + V_{pp\pi} \right\}.
\ee
Then, the energy spectrum at the $\Gamma$ point becomes
\be
E_{\Gamma_{VB}}^{\pm} = \epsilon_{\Gamma_{VB}} \pm D_\Gamma (1+\cos k_z c) \label{eq:E_Gamma}
\ee
for which the energy difference is the maximum at $k_z = 0$ and vanishes at the zone boundary.

If one experimentally measures the bandwidth at $\Gamma$ along the $k_z$ direction, one can extract the relation between two fundamental interlayer hopping parameters $V_{pp\sigma}$ and $V_{pp\pi}$ from the Eq. (3.10) and Eq. (3.11).  As an example, for MoS$_2$, we obtain $V_{pp\sigma} = 0.6344 $ and $V_{pp\pi} = -0.0592$ in eV, assuming $\delta = 3.4261$ (S-S distance) and $\delta_z = 2.9$.\cite{shiang,stewart}
As a result, we estimate $D_\Gamma = 0.4284$ eV, so that the bandwidth at $\Gamma$ point is about 1.7 eV which is comparable to the experimental results. These parameters are obtained from a simple exponential form of the overlap integral and may be tuned for realistic systems.

Now, we consider the VB and CB states at the $K$ point. For these states, the orbital composition is completely different from that of the states at the $\Gamma$ point. At K, states have equal contribution from $p_x$ and $p_y$ orbitals while $p_z$ orbital contribution is almost absent. The orbital composition is given by
\be
|\psi_{K_{VB}}\rangle = \tilde{c}_6 |d_2^{(e)}\rangle + c_6 |p_1^{(e)}\rangle \label{eq:orbital_K_val}
\ee
and
\be
|\psi_{K_{CB}}\rangle = \tilde{c}_5 |d_0^{(e)}\rangle + c_5 |p_{-1}^{(e)}\rangle \label{eq:orbital_K_cond}
\ee
where $|d_0^{(e)}\rangle = | d_{z^2}\rangle$, $|d_2^{(e)}\rangle = ( | d_{x^2-y^2}^{(e)}\rangle +i| d_{xy}^{(e)}\rangle ) /\sqrt{2}$ and $|p_{\pm 1}^{(e)}\rangle = (|p_x^{A}\rangle  + |p_x^{B}\rangle ) \pm i(|p_y^{A}\rangle  + |p_y^{B}\rangle )$. Here, $\tilde{c}_n = \sqrt{1-c_n^2}$.
In this case, we consider following effective 4$\times$4 Hamiltonian for CB and VB of the 3D bulk system.
\be
H_K \approx \bpm \epsilon_{K_{VB}} & 0 & \Delta_{K_{VB},k_z} & \alpha_{K,k_z} \\ 0 & \epsilon_{K_{CB}} & \beta_{K,k_z} & \Delta_{K_{CB},k_z} \\ \Delta^*_{K_{VB},k_z} & \beta^*_{K,k_z} & \epsilon_{K_{VB}} & 0 \\ \alpha^*_{K,k_z} & \Delta^*_{K_{CB},k_z} & 0 & \epsilon_{K_{CB}}\epm~\label{eq:eff_ham_K_0}
\ee
where $\Delta_{K_{VB},k_z}$ and $\Delta_{K_{CB},k_z}$ are mixing between same orbitals and $\alpha_{K,k_z}$ and $\beta_{K,k_z}$ are between different ones.
The upper (lower) $2\times 2$ diagonal block is for the upper (lower) slab.

\begin{figure}
\centering \epsfxsize=8.5cm \epsfbox{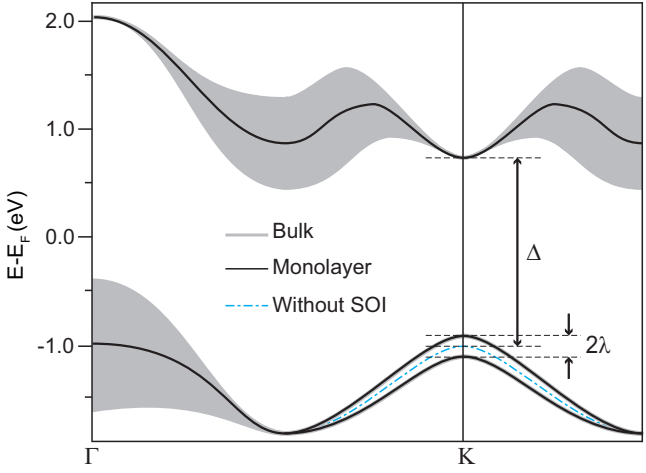}
\caption{Monolayer MX$_2$ bands (solid black line) and projected bulk bands (gray region) along the $\Gamma$ - $K$ direction, extracted from Ref. [35]. Blue dash-dot line is the band structure around the $K$ point when there is no SOI.}\label{fig3}
\end{figure}

As in the previous case, one can approximately evaluate $\Delta_{K_{VB},k_z}$ as follows. Neglecting terms involving $d$ orbitals, we have
\be
\Delta_{K_{VB},k_z} &=& \langle \Psi_{K_{VB},k_z}^{u}| H^\prime | \Psi_{K_{VB},k_z}^{l}\rangle \\
&=& \frac{c_6^2}{4}\sum_i \left\{ \left(V_{pp\sigma} - V_{pp\pi}\right)\frac{\delta_{i,\perp}^2}{\delta^2} +2V_{pp\pi}\right\}e^{i\v K\cdot\delta_i} \nn
&& + \frac{c_6^2}{4}\sum_i \left\{ \left(V_{pp\sigma} - V_{pp\pi}\right)\frac{\tilde{\delta}_{i,\perp}^2}{\tilde{\delta}^2} +2V_{pp\pi}\right\}\nn
&& \times e^{i\v K\cdot\tilde{\delta}_i}e^{ik_z c}
\ee
where $\delta_{i,\perp}^2 = \delta_{i,x}^2+\delta_{i,y}^2$ and $\v K$ represents the position of a $K$ point in the Brillouin zone of monolayer MX$_2$.
Since $\delta_{i,\perp}^2$ values for all the nearest neighbor hopping are the same, one can set $\delta_{\perp}^2=\delta_{i,\perp}^2$. Then $\Delta_{K_{VB},k_z}$ can be further simplified to
\be
\Delta_{K_{VB},k_z} = R_{K_{VB}}\left(f(\v K)+f(\v K)^* e^{ik_z c}\right)
\ee
where
\be
R_{K_{VB}} = \frac{c_6^2}{4}\left\{ \left(V_{pp\sigma} - V_{pp\pi}\right)\frac{\delta_{\perp}^2}{\delta^2} +2V_{pp\pi}\right\}
\ee
and
\be
f(\v k) = \sum_i e^{i\v k_{\perp}\cdot\delta_i}.
\ee
Refer to Appendix A for details. As was in the case of the graphene Dirac point, $f(\v K) = 0$ and we arrive at the conclusion that $\Delta_{K_{VB},k_z} = 0$ at the $K$ point\cite{neto}. By the same procedure, one can easily find that $\Delta_{K_{CB},k_z}$ is also vanishing. As a result, the matrix representation of the effective Hamiltonian reduces to
\be
H_K \approx \bpm \epsilon_{K_{VB}} & 0 & 0 & \alpha_{K,k_z} \\ 0 & \epsilon_{K_{CB}} & \beta_{K,k_z} & 0 \\ 0 & \beta^*_{K,k_z} & \epsilon_{K_{VB}} & 0 \\ \alpha^*_{K,k_z} & 0 & 0 & \epsilon_{K_{CB}}\epm.
\ee
$\alpha_{K,k_z}$ and $\beta_{K,k_z}$ in the Hamiltonian are calculated to be
\be
\alpha_{K,k_z} &=& \langle \Psi_{K_{VB},k_z}^{u}| H^\prime | \Psi_{K_{CB},k_z}^{l}\rangle \\
&\approx & D_K
\ee
and
\be
\beta_{K,k_z} &=& \langle \Psi_{K_{CB},k_z}^{u}| H^\prime | \Psi_{K_{VB},k_z}^{l}\rangle \\
&\approx & D_K e^{ik_z c}
\ee
where
\be
D_K = \frac{3c_5c_6}{4}\left(\frac{\delta_{\perp}}{\delta}\right)^2\left( V_{pp\sigma} - V_{pp\pi}\right).\quad
\ee
Note that $\alpha_{K,k_z} \neq \beta_{K,k_z}^*$ due to the layer index. Here, $\delta_{1,\perp} = (a/2,a/2\sqrt{3},0)$, $\delta_{2,\perp} = (-a/2,a/2\sqrt{3},0)$, and $\delta_{3,\perp} = (0,-a/\sqrt{3},0)$
(details are given in Appendix A). For the case of MoS$_2$ as an example, we have $D_K \approx 0.0263$ eV from the parameters given by $\delta_\perp = 1.8244 \angstrom$, $\delta = 3.4261 \angstrom$, $ V_{pp\sigma}=0.6344$eV, and $V_{pp\pi}=-0.0592$eV.\cite{shiang,stewart}

Finally, the effective Hamiltonian at the $K$ point becomes
\be
H_K \approx \bpm \epsilon_{K_{VB}} & 0 & 0 & D_K  \\ 0 & \epsilon_{K_{CB}} & D_K e^{ik_c} & 0 \\ 0 & D_K e^{-ik_c} & \epsilon_{K_{VB}} & 0 \\ D_K  & 0 & 0 & \epsilon_{K_{CB}}\epm.
\ee
Its eigenvalues are evaluated to be
\be
E_K^\pm = \frac{\epsilon_{VB} + \epsilon_{CB} \pm \sqrt{(\epsilon_{VB} - \epsilon_{CB})^2+4D_K^2}}{2}
\ee
which is independent of $k_z$.
Since $|\epsilon_{VB} - \epsilon_{CB}| \gg 2D_K$, as an approximation, we just have
\be
\epsilon_{K_{VB}} \rightarrow \epsilon_{K_{VB}} - \frac{D_K^2}{\epsilon_{CB} - \epsilon_{VB}} \label{eq:E_K_VB}
\ee
and
\be
\epsilon_{K_{CB}} \rightarrow \epsilon_{K_{CB}} + \frac{D_K^2}{\epsilon_{CB} - \epsilon_{VB}}.\label{eq:E_K_CB}
\ee
This means that the CB and VB energies at the $K$ point are $k_z$-independent and that they are subject to tiny energy shifts as we go from the monolayer to bulk cases.
In obtaining the results, two factors were crucial. First, there are no $p_z$ orbital components in both the CB and VB states at the $K$ point. Second, we have diminishing sums of phase factors due to the C3 symmetry.
One can obtain the same results for $K^\prime$ since the basis wave vectors are just complex conjugates of the wave vectors at $K$, namely, Eq. (\ref{eq:orbital_K_val}) and (\ref{eq:orbital_K_cond}).

\begin{figure*}
\centering \epsfxsize=12cm \epsfbox{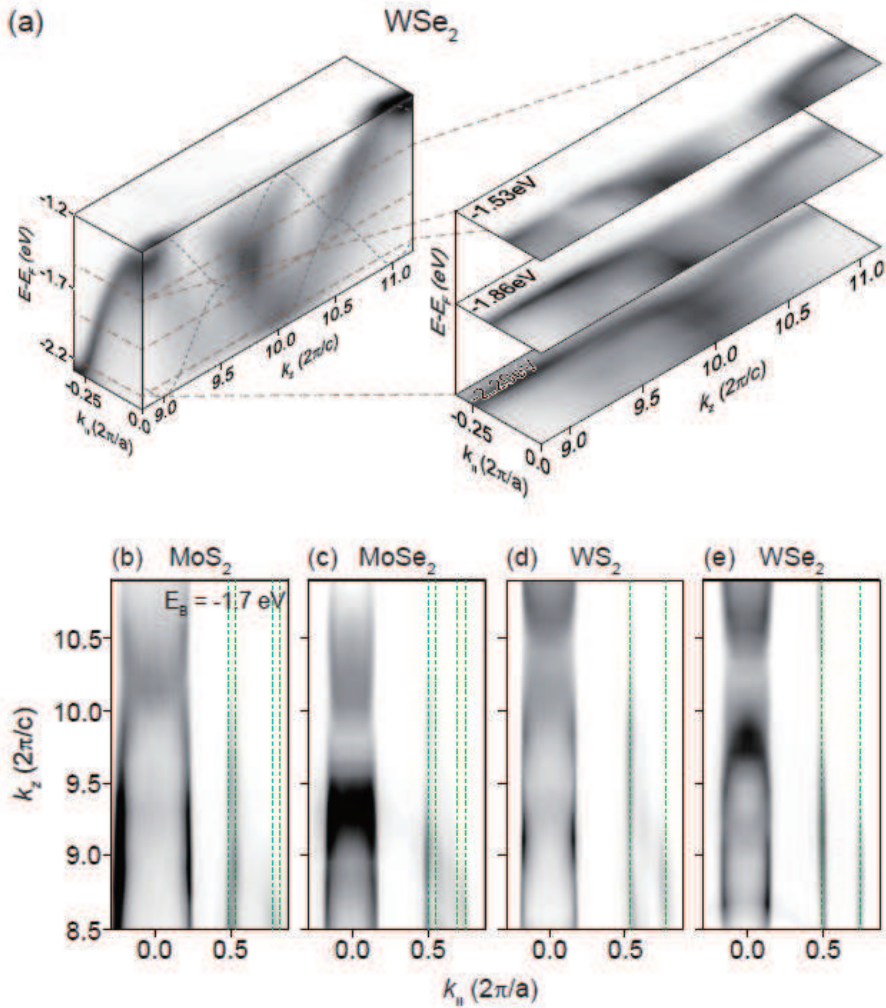}
\caption{(a) Intensity plot of WSe$_2$ ARPES data in energy and momentum ($k_z$, $k_\parallel$) space. $k_z$ dependent ARPES is taken by using different photon energies. $k_z$ of 9.0 and 11.0 correspond to the incident photon energies of 58 and 94 eV, respectively. The black dashed lines indicate the expected $k_z$ dispersion of the bands with $D$$_\Gamma$=0.3 eV [Eq. 3.11]. Three selected cuts on the right hand side along the brown dashed lines are ARPES intensity maps at constant energies in the momentum space ($k_z$, $k_\parallel$). Also shown are ARPES intensity maps of (b) MoS$_2$, (c) MoSe$_2$, (d) WS$_2$, (e) WSe$_2$ at a constant binding energy of $-1.7$ eV. The dashed lines are guides to eye for the electronic states near the $K$ point. These lines are straight along the $k$$_z$.}\label{fig4}
\end{figure*}

\begin{figure*}
\centering \epsfxsize=17cm \epsfbox{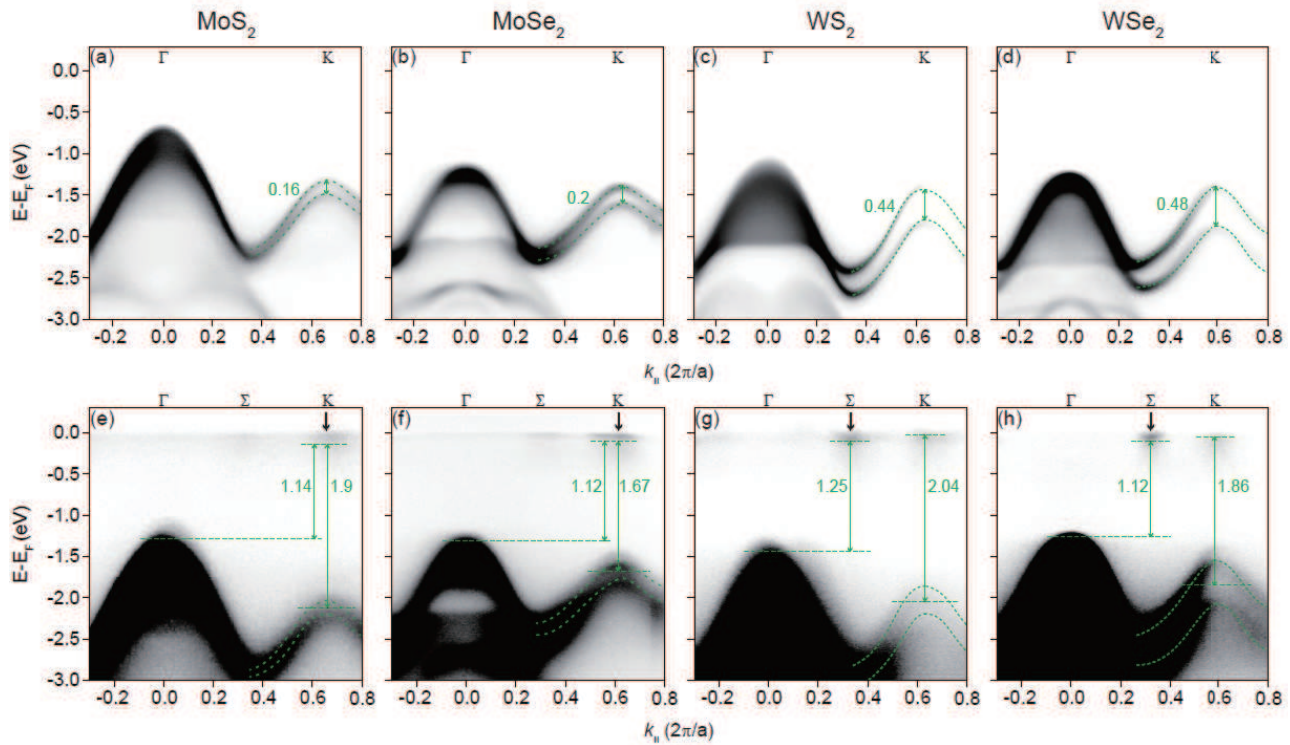}
\caption{(a)-(d) ARPES data along the $\Gamma$ to $K$ from MoS$_2$, MoSe$_2$, WS$_2$, and WSe$_2$. Green dotted lines indicate the band dispersions near the $K$ point. (e)-(h) ARPES data after potassium evaporation. The concentration of the doped electrons by potassium evaporation can be estimated from the Fermi surface volume. The estimated electron doping concentrations are $4.6\times 10^{13}$ $cm^{-2}$, $7.2\times 10^{13}$ $cm^{-2}$, $11\times 10^{13}$ $cm^{-2}$, and $7.6\times 10^{13}$ $cm^{-2}$ for MoS$_2$, MoSe$_2$, WS$_2$, and WSe$_2$, respectively.}\label{fig5}
\end{figure*}

Summarizing the theoretical part, upon stacking of MX$_2$ slabs, we find that the $k_z$ dispersions at two high symmetry points $\Gamma$ and $K$ are completely different. This can be interpreted in terms of the orbital composition and the discrete rotational symmetry of the system at those points. At the $\Gamma$ point, the eigenstates mainly consist of the out-of-plane orbitals such as the $d_{z^2}$ orbital of the M atom and the $p_z$ orbital of the X atom. As a result, the overlap integral between different layers is expected to be large compared to the in-plane orbitals. Since phase cancellations from the nearest neighbor hoppings are not possible at the $\Gamma$ point ($k_x=k_y=0$), the resulting energy spectra of the 3D MX$_2$ become dispersive along the $k_z$ direction.

At the $K$ point, on the other hand, we have both the out-of-plane ($d_{z^2}$) and in-plane orbital ($p_x$ and $p_y$) components for M and X atoms, respectively. Among them, only the $p_x$ and $p_y$ orbitals are responsible for the interlayer coupling because the distance between M atoms in the neighboring slabs is much larger than that of the nearest neighbor X atoms and thus the overlap between $d_{z^2}$ orbitals is negligible. One may immediately expect a small dispersion along the $k_z$ direction due to the small inter-plane hopping between $p_x$ and $p_y$ orbitals compared to the $p_z$ orbitals. However, we have shown that even this small dispersion is suppressed and the band becomes almost dispersionless along the $k_z$ direction due to the graphene-like phase cancellation among the nearest hopping processes stemming from the C3 symmetry of the system.

The experimental perspectives of the above results are as follows. The direct band gap at the K-point in MX$_2$ remains almost the same with the gap of the monolayer. Spin band splitting is expected to depend only on the atomic spin-orbit coupling of M atom in MX$_2$ and should be independent of the number of layers. These results tell us that we can extract the massive Dirac Fermion parameters from the electronic structure of MX$_2$. Figure 3 shows the expected ARPES data from MX$_2$ as ARPES captures a range of $k_z$ due to the finite photoelectron escape-depth.

\subsection{ARPES measurements on bulk 2H-MX$_2$}
As our theoretical work shows that we can extract the appropriate parameters from MX$_2$ data, we performed photon energy dependent ARPES to obtained the $k_z$ dispersive electronic structure. Figure 4(a) shows the ARPES data taken with incident photon energies between 50 and 100 eV near the in-plane $\Gamma$ point. Black dashed lines indicate band dispersions expected from Eq. (3.11). The data is in good agreement with the calculation results and shows a strong $k_z$ dispersion. The breadth in the ARPES data in the energy direction is due to the finite escape depth of the ARPES process (finite $k_z$ resolution). $k_z$ dispersions in MoS$_2$, MoSe$_2$, and WS$_2$ near the in-plane $\Gamma$ point are as strong as that in WSe$_2$ [Fig. 4(b),(c),(d)].

On the other hand, photon energy dependent ARPES data show no $k_z$ dispersion near the $K$ point as seen in Fig. 4(b)-(e), consistent with our calculation results in Eq. (3.27). Dashed lines in Fig. 4(b)-(e) are guides to eye which are straight (that is, no $k_z$ dispersion). Since the energy of the band at a specific in-plane momentum is the same regardless of $k_z$, ARPES spectra near the $K$ point are very sharp in comparison to the $\Gamma$ point data, both in the energy and in-plane momentum directions. This fact can be seen in Fig. 4(b)-(e) as well as in Fig. 5(a)-(d).

In order to extract the massive Dirac fermion parameters, we need ARPES data along the in-plane $\Gamma$ to $K$ (see Fig. 5). $2\lambda$ of MoS$_2$, MoSe$_2$, WS$_2$, and WSe$_2$ can be clearly observed in the data shown in Fig. 5(a)-(d). $2\lambda$ is drastically increased as the transition metal changes from Mo to W since $2\lambda$ mostly relies on the atomic spin-orbit coupling of the transition metal atom. The effective hopping integral, $t$, can also be estimated by fitting the data in Fig. 5(a)-(d) because $t$ is linearly proportional to the curvature of valence band dispersion at the $K$ point. $t$ of MoS$_2$ is, therefore, larger than that of MoSe$_2$ as the curvature is larger in MoS$_2$ than in MoSe$_2$ as can be seen from the data in Fig. 5(a),(b). The extracted $t$ values for MoS$_2$, MoSe$_2$, WS$_2$, and WSe$_2$ are given in table I.

In order to observe the direct band-gap size, $\Delta-\lambda$, at the $K$ point, it is necessary to see the bottom of the CB. The problem is that the states are not occupied and thus cannot be observed by ARPES. One way to circumvent the problem is to populate the CB bottom by potassium evaporation. Potassium has very low electron affinity and, when dosed on the sample surface, provides electrons. ARPES experiments after the potassium evaporation reveal the CB minimum (CBM) from which we can determine $\Delta$ [Fig. 5(e)-(h)]. The CBM is found to be located at the $K$ point in MoS$_2$ and MoSe$_2$, while it is located at the $\Sigma$ point in WS$_2$ and WSe$_2$. We note that the CBM of monolayer WS$_2$ and WSe$_2$ is located at the $K$ point. This is because the $k_z$ dispersion at the $\Sigma$ point for WS$_2$ and WSe$_2$ causes the CBM at the $\Sigma$ point to be located even lower than that at the $K$ point. Here, we emphasize that CB and VB near the $K$ point are not affected when layers are stacked and that, as a result, the massive Dirac fermion parameters including $\Delta$ could be correctly observed.

\begin{table}
\caption{\label{tab:table1} Parameters for the massive Dirac fermion model determined from the bulk ARPES data. Also given in the table are the values from published ARPES data on monolayers grown on various substrates. The parameters are expressed in unit of eV. Note that $t$ values with * mark are obtained by fitting the dispersions of the published data.}
\begin{ruledtabular}
\begin{tabular}{cccccccc}
 & $\Delta$ & 2$\lambda$ & $t$ & $\Delta$-$\lambda$\\
\hline
MoS$_2$ & 1.90 & 0.16 & 1.01 & 1.82\\
MoSe$_2$ & 1.67 & 0.20 & 0.90 & 1.57\\
WS$_2$ & 1.86 & 0.44 & 1.25 & 1.82\\
WSe$_2$ & 2.04 & 0.48 & 1.13 & 1.62\\
\hline
MoS$_2$/Au(111) [30] & 1.465 & 0.15 & 1.10* & 1.39\\
MoSe$_2$/bilayer graphene [5] & 1.67 & 0.18 & 0.90* & 1.58\\
WS$_2$/Au(111) [31] & & 0.42 & & \\
\end{tabular}
\end{ruledtabular}
\end{table}

All the parameters are summarized in the upper part of Table I. These values can be regarded as those for free standing monolayers. Using these experimentally obtained parameters, we sketch the expected minimal band structures of MoS$_2$, MoSe$_2$, WS$_2$, and WSe$_2$ near the K point [Fig. 6]. Note that the $t$ value affects the curvature of the VB and CB, and $\Delta-\lambda$ indicates the estimated direct band-gap size.

In the lower part of Table I, we also list the parameters determined from the published data. In comparing the values, one finds the values for free standing monolayer MoSe$_2$ (predicted) and monolayer MoSe$_2$ grown on bilayer-graphene are very similar. This can be attributed to the fact that the lattice mismatch between bilayer-graphene and MoSe$_2$ is only $\approx 0.3$ \%. On the other hand, free standing monolayer MoS$_2$ and monolayer MoS$_2$ on Au(111) have quite different parameters because of the large lattice mismatch between MoS$_2$ and the substrate. The latter case demonstrates the effect of the substrate on the electronic structures of the monolayer.


$t$ and 2$\lambda$ determined by ARPES are in quantitative agreements with the results from the first principles calculations\cite{Xiao}. Variation of $\Delta$ in MX$_2$ qualitatively agrees with the results from the first principles calculations, but $\Delta$ observed by ARPES is consistently 0.2 eV larger than that from the first principles calculations\cite{Xiao}. Since potassium evaporation concentration dependent experiments on WSe$_2$ recently revealed that $\Delta$ decreases from 1.6 to 1.45 eV \cite{King}, $\Delta$ of pristine MX$_2$ is expected to be even larger. Therefore, the first principles calculations on MX$_2$ clearly underestimate the true $\Delta$. This along with the substrate effect discussed above justify our study.

\begin{figure}
\centering \epsfxsize=8.5cm \epsfbox{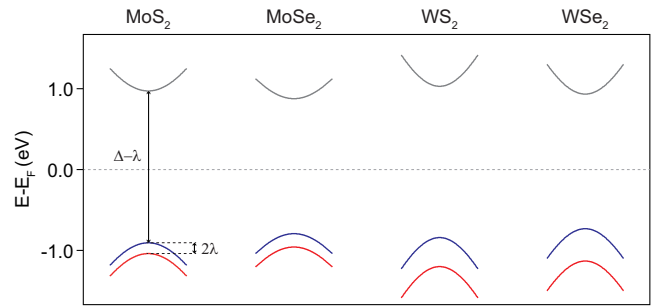}
\caption{The electronic structures of monolayer MoS$_2$, MoSe$_2$, WS$_2$, and WSe$_2$ near the K point predicted by the massive Dirac fermion model with the values in Table 1. Blue and red lines indicate the spin lifted valence band.}\label{fig6}
\end{figure}

In summary, we performed systematic ARPES studies on MX$_2$ (M = Mo, W; X = S, Se) family and determined the massive Dirac fermion parameters of monolayer MX$_2$ with the assistance from tight binding calculations which clearly shows how to determine the parameters from the electronic structure of MX$_2$. Our work provides the fundamental information on the quantitative understanding of the electrical and optical properties of this material family.

\acknowledgments
We thank Yeongkwan Kim, Jonathan D. Denlinger, Jongkeun Jung, and Soohyun Cho for assistance in the experiments. We also thank Wonshik Kyung for helpful discussions. This work was supported by the Incheon National University Research Grant in 20130817. B.S.K. and C.K. were supported by IBS-R009-D2, Korea.

\appendix
\section{Details on the tight binding analysis}
In this section, we show detailed derivation of the Hamiltonian matrix elements of the 3D bulk MX$_2$ systems at the $\Gamma$ and $K$ points.

The band splitting at $\Gamma$ point induced by the stacking of MX$_2$ layers is evaluated to be
\be
\Delta_{\Gamma,k_z} &=& \langle \Psi_{\Gamma_{VB},k_z}^{u}| H^\prime | \Psi_{\Gamma_{VB},k_z}^{l}\rangle  \label{eq:gamma_1} \\
&\approx & c_1^2 \langle p^{n,u}_z | H^\prime \sum_i | p^{n,l}_z(\delta_i) \rangle \nn
&& +c_1^2 \langle p^{n,u}_z | H^\prime \sum_i | p^{n+1,l}_z (\tilde{\delta}_i) \rangle e^{ik_z c} \quad \label{eq:gamma_2}\\
&=& \frac{c_1^2}{2}\Big( \langle p^{A,n,u}_z | - \langle p^{B,n,u}_z | \Big) H^\prime \sum_i \Big( |p^{A,n,l}_z(\delta_i) \rangle \nn
&& - p^{B,n,l}_z(\delta_i) \rangle \Big) +\frac{c_1^2}{2}\Big( \langle p^{A,n,u}_z | - \langle p^{B,n,u}_z | \Big) \label{eq:gamma_3} \\
&& \times H^\prime \sum_i \Big( |p^{A,n+1,l}_z(\tilde{\delta}_i)  \rangle -|p^{B,n+1,l}_z(\tilde{\delta}_i)  \rangle \Big)e^{ik_z c} \nn
&\approx & -\frac{c_1^2}{2}\langle p^{B,n,u}_z | H^\prime \sum_i |p^{A,n,l}_z(\delta_i)  \rangle \nn
&&-\frac{c_1^2}{2}\langle p^{A,n,u}_z | H^\prime \sum_i |p^{B,n+1,l}_z(\tilde{\delta}_i)\rangle e^{ik_z c}\label{eq:gamma_4}
\ee
where $\delta_i$ represents three nearest neighbor sites between two planes in the same unit cell, and $\tilde{\delta}_i = -\delta_i$ is for another pair of planes in different unit cells.
From (\ref{eq:gamma_1}) to (\ref{eq:gamma_2}), all overlap integrals between $d$ and $p$ are neglected. In addition, in obtaining (\ref{eq:gamma_4}), we assume that the nearest neighbor hoppings are dominant.

Derivation of the matrix elements of (\ref{eq:eff_ham_K_0}) is as follows.
First, $\Delta_{K_{VB},k_z}$ is
\begin{widetext}
\be
\Delta_{K_{VB},k_z} &=& \langle \Psi_{K_{VB},k_z}^{u}| H^\prime | \Psi_{K_{VB},k_z}^{l}\rangle  \label{eq:K_1} \\
&\approx & c_6^2 \langle p^{(e),n,u}_1 | H^\prime \sum_i | p^{(e),n,l}_1(\delta_i) \rangle e^{i\v K\cdot\delta_i} +c_6^2 \langle p^{(e),n,u}_1 | H^\prime \sum_i | p^{(e),n+1,l}_1(\tilde{\delta}_i) \rangle e^{i\v K\cdot\tilde{\delta}_i} e^{ik_z c} \quad \label{eq:K_2}\\
&=& \frac{c_6^2}{4}\sum_i \left\{ \left(\langle p_x^{A,n,u}| +\langle p_x^{B,n,u}|\right) -i\left( \langle p_y^{A,n,u}| +\langle p_y^{B,n,u}|\right) \right\}H^\prime \big\{ \left(|p_x^{A,n,l}(\delta_i)\rangle +|p_x^{B,n,l}(\delta_i)\rangle \right) \nn
&& +i\big(|p_y^{A,n,l}(\delta_i)\rangle +|p_y^{B,n,l}(\delta_i)\rangle \big)\big\}e^{i\v K\cdot\delta_i} +\frac{c_6^2}{4}\sum_i \left\{ \left(\langle p_x^{A,n,u}| +\langle p_x^{B,n,u}|\right) -i\left( \langle p_y^{A,n,u}| +\langle p_y^{B,n,u}|\right) \right\}H^\prime \nn
&& \times \big\{ \left(|p_x^{A,n+1,l}(\tilde{\delta}_i)\rangle +|p_x^{B,n+1,l}(\tilde{\delta}_i)\rangle \right) +i\big(|p_y^{A,n+1,l}(\tilde{\delta}_i)\rangle +|p_y^{B,n+1,l}(\tilde{\delta}_i)\rangle \big)\big\}e^{i\v K\cdot\tilde{\delta}_i}e^{ik_z c} \\
&\approx & \frac{c_6^2}{4}\sum_i \left( \langle p_x^{B,n,u}| -i\langle p_y^{B,n,u}|\right) H^\prime \left( |p_x^{A,n,l}(\delta_i)\rangle +i|p_y^{A,n,l}(\delta_i)\rangle  \right)e^{i\v K\cdot\delta_i} \nn
&& +\frac{c_6^2}{4}\sum_i \left(  \langle p_x^{A,n,u}| -i\langle p_y^{A,n,u}| \right)H^\prime \left( |p_x^{B,n,l}(\tilde{\delta}_i)\rangle +i|p_y^{B,n,l}(\tilde{\delta}_i)\rangle \right)e^{i\v K\cdot\tilde{\delta}_i}e^{ik_z c} \label{eq:K_3}
\ee
\end{widetext}
where $\v K$ is the position of a $K$ point in the momentum space of MX$_2$ monolayer. We also retained only the dominant nearest neighbor hoppings in obtaining (\ref{eq:K_2}) and (\ref{eq:K_3}).
Since $t^{(LL)}_{p^\prime_i p_j}$ is invariant under $i\leftrightarrow j$, the terms involving $p_x$ and $p_y$ simultaneously cancel each other. As a result, $\Delta_{K_{VB},k_z}$ becomes
\begin{widetext}
\be
\Delta_{K_{VB},k_z} &=& \frac{c_6^2}{4}\sum_i \left( \langle p_x^{B,n,u}|H^\prime |p_x^{A,n,l}(\delta_i)\rangle +\langle p_y^{B,n,u}| H^\prime |p_y^{A,n,l}(\delta_i)\rangle  \right)e^{i\v K\cdot\delta_i} \nn
&& +\frac{c_6^2}{4}\sum_i \left(  \langle p_x^{A,n,u}|H^\prime |p_x^{B,n,l}(\tilde{\delta}_i)\rangle +\langle p_y^{A,n,u}|H^\prime |p_y^{B,n,l}(\tilde{\delta}_i)\rangle \right)e^{i\v K\cdot\tilde{\delta}_i}e^{ik_z c} \\
&=& \frac{c_6^2}{4}\sum_i \left\{ \left(V_{pp\sigma} - V_{pp\pi}\right)\frac{\delta_{i,\perp}^2}{\delta^2} +2V_{pp\pi}\right\}e^{i\v K\cdot\delta_i} + \frac{c_6^2}{4}\sum_i \left\{ \left(V_{pp\sigma} - V_{pp\pi}\right)\frac{\tilde{\delta}_{i,\perp}^2}{\tilde{\delta}^2} +2V_{pp\pi}\right\}e^{i\v K\cdot\tilde{\delta}_i}e^{ik_z c} \quad~ \\
&=& \frac{c_6^2}{4}\sum_i \left\{ \left(V_{pp\sigma} - V_{pp\pi}\right)\frac{\delta_{i,\perp}^2}{\delta^2} +2V_{pp\pi}\right\}\left(e^{i\v K\cdot\delta_i} + e^{i\v K\cdot\tilde{\delta}_i}e^{ik_z c}\right) \\
&=& \frac{c_6^2}{4} \left\{ \left(V_{pp\sigma} - V_{pp\pi}\right)\frac{\delta_{\perp}^2}{\delta^2} +2V_{pp\pi}\right\} \sum_i \left(e^{i\v K\cdot\delta_i} + e^{-i\v K\cdot\delta_i}e^{ik_z c}\right)
\ee
\end{widetext}
where we use the fact that $\delta_{\perp}^2 \equiv \delta_{i,\perp}^2 = \delta_{i,x}^2+\delta_{i,y}^2$ is independent of $i$. With $\delta_{1,\perp} = (a/2,a/2\sqrt{3},0)$, $\delta_{2,\perp} = (-a/2,a/2\sqrt{3},0)$, and $\delta_{3,\perp} = (0,-a/\sqrt{3},0)$, one can show that $\sum_i e^{i\v K\cdot\delta_i} =\sum_i e^{-i\v K\cdot\delta_i} =0$.

The other matrix elements related to the slight shifts of the band edges are evaluated as follows.
\begin{widetext}
\be
\alpha_{K,k_z} &=& \langle \Psi_{K_{VB},k_z}^{u}| H^\prime | \Psi_{K_{CB},k_z}^{l}\rangle \\
&\approx & c_5 c_6 \langle p^{(e),n,u}_1 | H^\prime \sum_i | p^{(e),n,l}_{-1}(\delta_i) \rangle e^{i\v K\cdot\delta_i} +c_5 c_6\langle p^{(e),n,u}_1 | H^\prime \sum_i | p^{(e),n+1,l}_{-1}(\tilde{\delta}_i) \rangle e^{i\v K\cdot\tilde{\delta}_i} e^{ik_z c} \\
&\approx & -\frac{c_5 c_6 }{4}\sum_i \left( \langle p_x^{B,n,u}| +i\langle p_y^{B,n,u}|\right) H^\prime \left( |p_x^{A,n,l}(\delta_i)\rangle +i|p_y^{A,n,l}(\delta_i)\rangle  \right)e^{i\v K\cdot\delta_i} \nn
&& -\frac{c_5 c_6 }{4}\sum_i \left(  \langle p_x^{A,n,u}| +i\langle p_y^{A,n,u}| \right)H^\prime \left( |p_x^{B,n,l}(\tilde{\delta}_i)\rangle +i|p_y^{B,n,l}(\tilde{\delta}_i)\rangle \right)e^{i\v K\cdot\tilde{\delta}_i}e^{ik_z c} \\
&=& -\frac{c_5 c_6 }{4}\sum_i \left( \langle p_x^{B,n,u}|H^\prime |p_x^{A,n,l}(\delta_i)\rangle +\langle p_y^{B,n,u}| H^\prime |p_y^{A,n,l}(\delta_i)\rangle + 2i\langle p_x^{B,n,u}|H^\prime |p_y^{A,n,l}(\delta_i)\rangle \right)e^{i\v K\cdot\delta_i} \nn
&& -\frac{c_5 c_6 }{4}\sum_i \left(  \langle p_x^{A,n,u}|H^\prime |p_x^{B,n,l}(\tilde{\delta}_i)\rangle +\langle p_y^{A,n,u}|H^\prime |p_y^{B,n,l}(\tilde{\delta}_i)\rangle +2i\langle p_x^{B,n,u}|H^\prime |p_y^{A,n,l}(\tilde{\delta}_i)\rangle \right)e^{i\v K\cdot\tilde{\delta}_i}e^{ik_z c} \\
&=& -\frac{c_5 c_6 }{4}\sum_i \left\{ \left( V_{pp\sigma} - V_{pp\pi}\right)\frac{(\delta_{i,x} + i\delta_{i,y})^2}{\delta_i^2} +2V_{pp\pi}\right\}e^{i\v K\cdot\delta_i} \nn
&& -\frac{c_5 c_6 }{4}\sum_i\left\{ \left( V_{pp\sigma} - V_{pp\pi}\right)\frac{(\tilde{\delta}_{i,x} + i\tilde{\delta}_{i,y})^2}{\tilde{\delta}_i^2} +2V_{pp\pi}\right\}e^{i\v K\cdot\tilde{\delta}_i}e^{ik_z c} \\
&=& -\frac{c_5 c_6 }{4} \left( V_{pp\sigma} - V_{pp\pi}\right)\left(\frac{\delta_{\perp}}{\delta}\right)^2 \sum_i \left\{ \frac{(\delta_{i,x} + i\delta_{i,y})^2}{\delta_{\perp}^2} e^{i\v K\cdot\delta_i} + \frac{(\tilde{\delta}_{i,x} + i\tilde{\delta}_{i,y})^2}{\tilde{\delta}_{\perp}^2} e^{i\v K\cdot\tilde{\delta}_i}e^{ik_z c} \right\}
\ee
\end{widetext}
and, in the same way,
\begin{widetext}
\be
\beta_{K,k_z} &=& \langle \Psi_{K_{CB},k_z}^{u}| H^\prime | \Psi_{K_{VB},k_z}^{l}\rangle \\
&\approx & -\frac{c_5 c_6 }{4} \left( V_{pp\sigma} - V_{pp\pi}\right)\left(\frac{\delta_{\perp}}{\delta}\right)^2 \sum_i \left\{ \frac{(\delta_{i,x} - i\delta_{i,y})^2}{\delta_{\perp}^2} e^{i\v K\cdot\delta_i} + \frac{(\tilde{\delta}_{i,x} - i\tilde{\delta}_{i,y})^2}{\tilde{\delta}_{\perp}^2} e^{i\v K\cdot\tilde{\delta}_i}e^{ik_z c} \right\}.
\ee
\end{widetext}

At $K = (4\pi/3a,0,0)$, one can show that
\be
-3 &=& \sum_i \frac{(\delta_{i,x} + i\delta_{i,y})^2}{\delta_{\perp}^2} e^{i\v K\cdot\delta_i} = \sum_i\frac{(\tilde{\delta}_{i,x} - i\tilde{\delta}_{i,y})^2}{\tilde{\delta}_{\perp}^2} e^{i\v K\cdot\tilde{\delta}_i}\nn
\ee
and
\be
0 &=& \sum_i \frac{(\tilde{\delta}_{i,x} + i\tilde{\delta}_{i,y})^2}{\tilde{\delta}_{\perp}^2} e^{i\v K\cdot\tilde{\delta}_i} = \sum_i \frac{(\delta_{i,x} - i\delta_{i,y})^2}{\delta_{\perp}^2} e^{i\v K\cdot\delta_i}. \nn
\ee
As a result, we have simple formulae for $\alpha_{K,k_z}$ and $\beta_{K,k_z}$ as
\be
\alpha_{K,k_z} = \frac{3c_5c_6}{4}\left(\frac{\delta_{\perp}}{\delta}\right)^2\left( V_{pp\sigma} - V_{pp\pi}\right)
\ee
and
\be
\beta_{K,k_z} = \frac{3c_5c_6}{4}\left(\frac{\delta_{\perp}}{\delta}\right)^2\left( V_{pp\sigma} - V_{pp\pi}\right)e^{ik_z c}.
\ee
One can find that we get the same result at the $K^\prime$ point, $(-4\pi/3a,0,0)$.


\end{document}